\documentclass[a4paper]{jpconf-kb}
\usepackage{amsmath}
\usepackage{iopams}
\usepackage{natbib}
\usepackage{graphicx}
\usepackage{url}

\usepackage{imakeidx}
\makeindex[name=subject,title=Subject index,columns=1]
\makeindex[name=authors,title=Author index,columns=1]

\begin{document}
\graphicspath{{FIGURES/}}

\title{Torus quantum vortex knots in the Gross-Pitaevskii model for Bose-Einstein  condensates}

\author{D.\ Proment$^1$, M.\ Onorato$^{2,3}$ \& C.F.\ Barenghi$^4$}

\address{$^1$ School of Mathematics, University of East Anglia, Norwich Research Park, Norwich, NR4 7RJ, United Kingdom}
\address{$^2$ Dipartimento di Fisica, Universit{\`a} degli Studi di Torino, Via Pietro Giuria 1, 10125 Torino, Italy}
\address{$^3$ INFN, Sezione di Torino, Via Pietro Giuria 1, 10125 Torino, Italy}
\address{$^4$ School of Mathematics and Statistics, Newcastle University, Newcastle upon Tyne, NE1 7RU, United Kingdom}

\ead{davideproment@gmail.com}

\index[authors]{Proment, D.} \index[authors]{Onorato, M.} \index[authors]{Barenghi, C.F.}

\begin{abstract}
%
%
We examine on the static and dynamical properties of quantum knots in a Bose-Einstein condensate.
In particular, we consider the Gross-Pitaevskii model and revise a technique to construct {\it ab initio} the condensate wave-function of a generic torus knot.
After analysing its excitation energy, we study its dynamics relating the topological parameter to its translational velocity and characteristic size.
We also investigate the breaking mechanisms of non shape-preserving torus knots confirming an evidence of universal decaying behaviour previously observed.
\end{abstract}

\section{Introduction}
%
\index[subject]{Bose-Einstein condensate}
A Bose-Einstein condensate (BEC) is a peculiar state of matter that manifests macroscopically the effects of quantum mechanics. 
Essentially, when a system of weakly interacting bosons is cooled to very low temperatures, the great majority of them occupies the single particle {\it ground state}, i.e. the lowest energy quantum mechanical level. 
Then, an {\it order parameter} complex field $ \psi $, usually called condensate wave-function, can be used to describe the collective behaviour of all bosons.
Predicted by Bose and Einstein in the 1920s, BECs were first created experimentally by cooling specially confined alkali atom gases in 1995 [\cite{anderson1995observation, davis1995bose}].
Nowadays, a condensate can be obtained using also many other systems of interacting bosons like magnons [\cite{demokritov2006bose}], exciton polaritons [\cite{kasprzak2006bose}], and photons [\cite{klaers2010bose}].
\index[authors]{Anderson, M.H.}
\index[authors]{Ensher, J.R.}
\index[authors]{Matthews, M.R.}
\index[authors]{Wieman, C.E.}
\index[authors]{Cornell, E.A.}

\index[authors]{Davis, K.B.}
\index[authors]{Mewes, M.O.}
\index[authors]{Andrews, M.V}
\index[authors]{Van Druten, N.J.}
\index[authors]{Durfee, D.S.}
\index[authors]{Kurn, D.M.}
\index[authors]{Ketterle, W.}

\index[authors]{Demokritov, S.O.}
\index[authors]{Demidov, V.E.}
\index[authors]{Dzyapko, O.}
\index[authors]{Melkov, G.A.}
\index[authors]{Serga, A.A.}
\index[authors]{Hillebrands, B.}
\index[authors]{Slavin, A.N.}

\index[authors]{Kasprzak, J.}
\index[authors]{Richard, M.}
\index[authors]{Kundermann, S.}
\index[authors]{Baas, A.}
\index[authors]{Jeambrun, P.}
\index[authors]{Keeling, J.M.J.}
\index[authors]{Marchetti, F.M.}
\index[authors]{Szyma{\'n}ska, M.H.}
\index[authors]{Andr{\'e}, R.}
\index[authors]{Staehli, J.L.}
\index[authors]{Savona, V.}
\index[authors]{Littlewood, P.B.}
\index[authors]{Deveaud, B.}
\index[authors]{Le Si Dang}

\index[authors]{Klaers, J.}
\index[authors]{Schmitt, J.}
\index[authors]{Vewinger, F.}
\index[authors]{Weitz, M.}

\index[authors]{Sun, C.}
\index[authors]{Jia, S.}
\index[authors]{Barsi, C.}
\index[authors]{Rica, S.}
\index[authors]{Picozzi, A.}
\index[authors]{Fleischer, J.W.}

Among many models used to describe the dynamics of BECs, without any doubt the most simple one is the Gross-Pitaevskii equation (GPE) [\cite{pitaevskii2003bose}]
\index[authors]{Pitaevskii, L.P.}
\index[authors]{Stringari, S.}
\index[subject]{Gross-Pitaevskii model}
\begin{equation}
i \hbar \, \frac{\partial \psi(\mathbf{r}, t)}{\partial t} = -\frac{\hbar^2}{2 m} \nabla^2 \psi(\mathbf{r}, t) 
+ \frac{4\pi \hbar^2 a_s}{m} |\psi(\mathbf{r}, t)|^2 \psi(\mathbf{r}, t) + V_{ext}(\mathbf{r}, t) \, \psi(\mathbf{r}, t) \, ,
\label{eq:GPE}
\end{equation}
where $ \psi $ denotes the condensate wave-function in space and time, $ m $ and $ a_s $ are the mass and the $ s $-wave scattering length of the boson respectively, $ V_{ext} $ is an external potential acting on the condensate, and $ \hbar $ is the reduced Planck constant.
This nonlinear partial differential equation, also known in other research fields as the nonlinear Schr{\"o}dinger equation, mimics well the BEC dynamics in the case of a very dilute gas of bosons in the limit of zero temperature, that is to say when almost all bosons are in the ground state.
It is well known that the GPE is not integrable in more than one spatial dimension and so no generic analytical solutions may be provided.
\index[subject]{nonlinear Schr{\"o}dinger equation}

For simplicity we will consider in the following no external potential acting on the system, i.e. $ V(\mathbf{r}, t)_{ext} \equiv 0 $, and the same non-dimensional version of GPE chosen in \cite{berloff:2004}.
By rescaling the wave-function using the uniform density at infinity as $ \psi \rightarrow \sqrt{\rho_{\infty}} \, \psi $, time as $ t \rightarrow m/(8\pi\hbar  \, a_s \, \rho_{\infty}) \, t $, and lengths as $ \mathbf{r} \rightarrow \xi \, \mathbf{r} $, equation (\ref{eq:GPE}) results in
\index[authors]{Berloff, N.}
\begin{equation}
i \, \frac{\partial\psi(\mathbf{r}, t)}{\partial t} + \frac{1}{2} \nabla^2 \psi(\mathbf{r}, t) 
- \frac{1}{2} |\psi(\mathbf{r}, t)|^2 \psi(\mathbf{r}, t) = 0 \, ,
\quad \mbox{with} \quad \xi=\sqrt{\frac{1}{8 \pi \, a_s \, \rho_{\infty}}} \, .
\label{eq:ndGPE}
\end{equation}
We underline that $ \xi $, usually called the {\it healing length} or {\it coherence length}, is the only characteristic length of the system. 
It turns out to be only a function of the boson scattering length and the uniform density at infinity of the BEC, or equivalently the total number of bosons per the total volume when only few localised fluctuations on a uniform solution are present.

\index[subject]{superfluid}
It is straightforward to prove that a BEC modelled by the GPE is also a superfluid, that is to say it is a fluid having no viscosity. 
Indeed, one may use the Madelung transformation $ \psi(\mathbf{r}, t)= \sqrt{\rho(\mathbf{r}, t)} \, \exp{ i \theta(\mathbf{r}, t)} $ to rewrite the complex order parameter wave-function in terms of two real fields $ \rho(\mathbf{r}, t) $ and $ \theta(\mathbf{r}, t) $.
Substituting this into equation (\ref{eq:ndGPE}), splitting real and imaginary parts, and defining the vector field $ \mathbf{v}(\mathbf{r}, t)=\nabla \theta(\mathbf{r}, t) $, one obtains the following system of equations:
\begin{equation}
\begin{split}
& \frac{\partial \rho}{\partial t} + \nabla \cdot \left( \rho \, \mathbf{v} \right) = 0 \, , \\
& \frac{\partial \mathbf{v}}{\partial t} + \left(\mathbf{v} \cdot \nabla \right) \mathbf{v} = - \frac{\nabla \rho}{2} + \nabla \left( \frac{\nabla^2 \sqrt{\rho}}{2\sqrt{\rho}} \right) \, .
\end{split}
\end{equation}  
These are nothing but the continuity and linear momentum conservation equations for a barotropic, compressible, inviscid, and irrotational fluid where the two fields $ \rho(\mathbf{r}, t) $ and $ \mathbf{v}(\mathbf{r}, t) $ are simply the fluid density and velocity.
The last term in the latter equation is called the {\it quantum stress tensor}: it becomes important at scales of the order of the healing length and is the key difference with the classical Euler equation. 
\index[subject]{irrotational velocity field}

\index[subject]{superfluid}
\index[subject]{irrotational velocity field}
Even though the superfluid BEC is irrotational, vortical structures called quantized vortices can appear as the wave-function $ \psi $ is still single-valued if the phase field $ \theta $ varies by a multiple of $ 2\pi $.
Indeed, the circulation $ \mathcal{C} $ around the closed curve $ \gamma $ results in
\begin{equation}
\mathcal{C}=\oint_\gamma \mathbf{v} \cdot d\mathbf{l} = \oint_\gamma \nabla \theta \cdot d\mathbf{l} = \Delta \theta = \pm n \, 2\pi \, , \quad \mbox{with} \quad n \in \mathcal{N} \, .
\end{equation}
If $ n \neq 0 $, the circulation takes non-zero quantized values $ \pm n \, \kappa $, being $ \kappa=2\pi $ the quantum of circulation, and for Stokes' theorem the domain inside $ \gamma $ is no more simply connected.
In two-dimensional slices of the three-dimensional space these quantized vortices take the form of zeros of the density field (the density goes regularly to zero in the vicinity of the defect) where the phase twists by $ \pm \kappa $. 
In three dimensions the quantized vortices form closed rings or lines that end at the system boundaries.
The vortex lines interact with each other similarly to steady current-carrying wires, may form helicoidal Kelvin waves and can eventually reconnect rearranging their topology.
\index[subject]{Kelvin waves}

\index[authors]{Matthews, M.R.}
\index[authors]{Anderson, B.P.}
\index[authors]{Haljan, P.C.}
\index[authors]{Hall, D.S.}
\index[authors]{Wieman, C.E.}
\index[authors]{Cornell, E.A.}
\index[authors]{Madison, K.W.}
\index[authors]{Chevy, F.}
\index[authors]{Wohlleben, W.}
\index[authors]{Dalibard, J.}
\index[authors]{Abo-Shaeer, J.R.}
\index[authors]{Raman, C.}
\index[authors]{Vogels, J.M.}
\index[authors]{Ketterle, W.}
\index[authors]{Weiler, C.N.}
\index[authors]{Neely, T.W.}
\index[authors]{Scherer, D.R.}
\index[authors]{Bradley, A.S.}
\index[authors]{Davis, M.J.}
\index[authors]{Anderson, B.P.}
\index[authors]{Neely, T.W.}
\index[authors]{Samson, E.C.}
\index[authors]{Bradley, A.S.}
\index[authors]{Davis, M.J.}
\index[authors]{Anderson, B.P.}
\index[authors]{Anderson, B.P.}
\index[authors]{Haljan, P.C.}
\index[authors]{Regal, C.A.}
\index[authors]{Feder, D.L.}
\index[authors]{Collins, L.A.}
\index[authors]{Clark, C.W.}
\index[authors]{Cornell, E.A.}
Vortex lines have been widely observed in BECs forming arrays under rotation [\cite{PhysRevLett.83.2498, PhysRevLett.84.806, Abo-Shaeer20042001} or interacting and decaying in quasi two-dimensional geometries [\cite{Weiler:2008fk, PhysRevLett.104.160401}].
Vortex ring observation has been on the contrary quite elusive so far [\cite{PhysRevLett.86.2926}] as the visualisation techniques used in BECs do not allow for a clear measure of the density and velocity fields and in most cases are destructive measurements.      
Quantum vortices have also been experimentally observed in superlfuid Helium II [\cite{donnelly1991}] and even reconnection events have been directly detected using hydrogen particle tracers [\cite{bewley2008characterization}].
\index[authors]{Donnelly, R.J.}
\index[authors]{Bewley, G.P.}
\index[authors]{Paoletti, M.S.}
\index[authors]{Sreenivasan, K.R.}
\index[authors]{Lathrop, D.P.}
\index[subject]{reconnection}

\index[subject]{superfluid}
\index[subject]{Biot-Savart model}
\index[subject]{local induction approximation}
\index[subject]{stability}
More topologically complex quantum vortices like knots have not yet been detected in a superfluid.
Several theoretical and numerical studies have been carried out to understand the stability and dynamics of knots using the Biot-Savart model and its local induction approximation (LIA) [\cite{donnelly1991}].
Only recently we have been able to create an {\it ab initio} wave-function describing the simpler vortex knots in the GPE [\cite{PhysRevE.85.036306}].
This provided a basis for studying the shape preserving properties and dynamics of topological structures in BECs.
Very similar results regarding trefoil knot dynamics and breaking mechanisms were observed experimentally in classical fluid where a specifically-designed airfoil was created to produce the vortex knot [\cite{kleckner2013creation}].
\index[authors]{Donnelly, R.J.}
\index[authors]{Proment, D.}
\index[authors]{Onorato, M.}
\index[authors]{Barenghi, C.F.}
\index[authors]{Kleckner, D.}
\index[authors]{Irvine, W.T.M.}
\index[subject]{breaking mechanisms}
\index[subject]{trefoil knot}

In this contribution we extend our previous work on knots in the GPE, analysing the energy of simpler torus knots, reporting new results on more topologically complicated knots and presenting open questions and future perspectives in this field.
\index[authors]{Proment, D.}
\index[authors]{Onorato, M.}
\index[authors]{Barenghi, C.F.}

\section{The generic torus vortex knot $ T_{p, q} $ wave-function}
In \cite{PhysRevE.85.036306} we proposed a technique to generate the initial wave-function of a trefoil and its dual knot based on the fact that several knots may be built as closed lines on a torus.
\index[authors]{Proment, D.}
\index[authors]{Onorato, M.}
\index[authors]{Barenghi, C.F.}
Indeed, by twisting a curve $ p $ and $ q $ times around the toroidal and poloidal circles of the torus before tying it the line, one can build topologically non-trivial curves defined as $ \mathcal{T}_{p, q} $.
The easier example is the trefoil knot which is the $ \mathcal{T}_{2, 3} $ knot and its dual, the $ \mathcal{T}_{3, 2} $ knot.
\index[subject]{trefoil knot}

We may ask ``how is it possible to create a quantum vortex knot $ \mathcal{T}_{p, q} $ on a torus axisymmetric with respect to the $ z $ axis set at the origin of a cartesian three-dimensional space?''.
The main idea is to notice that each plane $ sOz $ perpendicular to the $ z $ axis of the torus identifies a particular slice of the three-dimensional space where $ 2q $ two-dimensional quantum vortices are present.
In particular, $ q $ vortices lie on the circumference of radius $ R_1 $ centered at $ (-R_0, 0) $ and the other $ q $ vortices lie on the circumference of radius $ R_1 $ centered at $ (R_0, 0) $, given $ R_0 $ and $ R_1 $ the toroidal and poloidal radii of the torus respectively.

In \cite{berloff:2004} a Pad{\'e} approximation for a two-dimensional wave-function $ \Psi_{2D}(s-s_0, z-z_0) $ describing a single quantum vortex centered at $ (s_0, z_0) $ is given; moreover, to create a quantum vortex with opposite circulation it is sufficient to apply the complex conjugate operator $ (\cdot)^\ast $ to the wave-function. 
\index[authors]{Berloff, N.}
To the first approximation, one can consider the two-dimensional wave-function describing $ 2q $ quantum vortices just as a superposition of the single vortex solutions, that is to say the multiplication of the single vortex wave-functions $ \Psi_{2D} $.
Thus, one can come back to the three-dimensional space knowing the map $ s(x, y) $ and controlling the position of the vortices in each slice $ sOz $ so that the three-dimensional quantum vortex line tie into the desired $ \mathcal{T}_{p, q} $ knot.
More specifically, the three-dimensional wave-function is given by
\begin{equation}
\begin{split}
\psi_{p, q}(x, y, z) = \, &
\prod_{i=1}^{q} \Psi_{2D}\left\{ s(x, y) - R_0 - R_1 \cos \left[\alpha(x, y) + i \frac{2\pi \,p}{q}\right], z - R_1 \sin \left[\alpha(x, y) + i \frac{2\pi \,p}{q}\right] \right\}
\\ \times \, &
\prod_{i=1}^{q} \Psi_{2D}^\ast \left\{ s(x, y) + R_0 + R_1 \cos \left[\alpha(x, y) + i \frac{2\pi \,p}{q}\right], z - R_1 \sin \left[\alpha(x, y) + i \frac{2\pi \,p}{q}\right] \right\},
\end{split}
\label{eq:Tpq}
\end{equation}
where
\begin{equation}
s(x, y)=\mbox{sgn}(x) \, \sqrt{x^2+y^2} \quad \mbox{and} \quad \alpha(x, y)=\frac{q \, \mbox{atan2}(x, y)}{p} \, .
\end{equation}
Here $ \mbox{sgn}(\cdot) $ is the sign function and $ \mbox{atan2}(\cdot, \cdot) $ is the arctangent function with two argument that gather the information on the signs of the inputs in order to return the appropriate quadrant of the computed angle. 

\begin{figure}[!ht]
\centering
\includegraphics[width=0.85\linewidth]{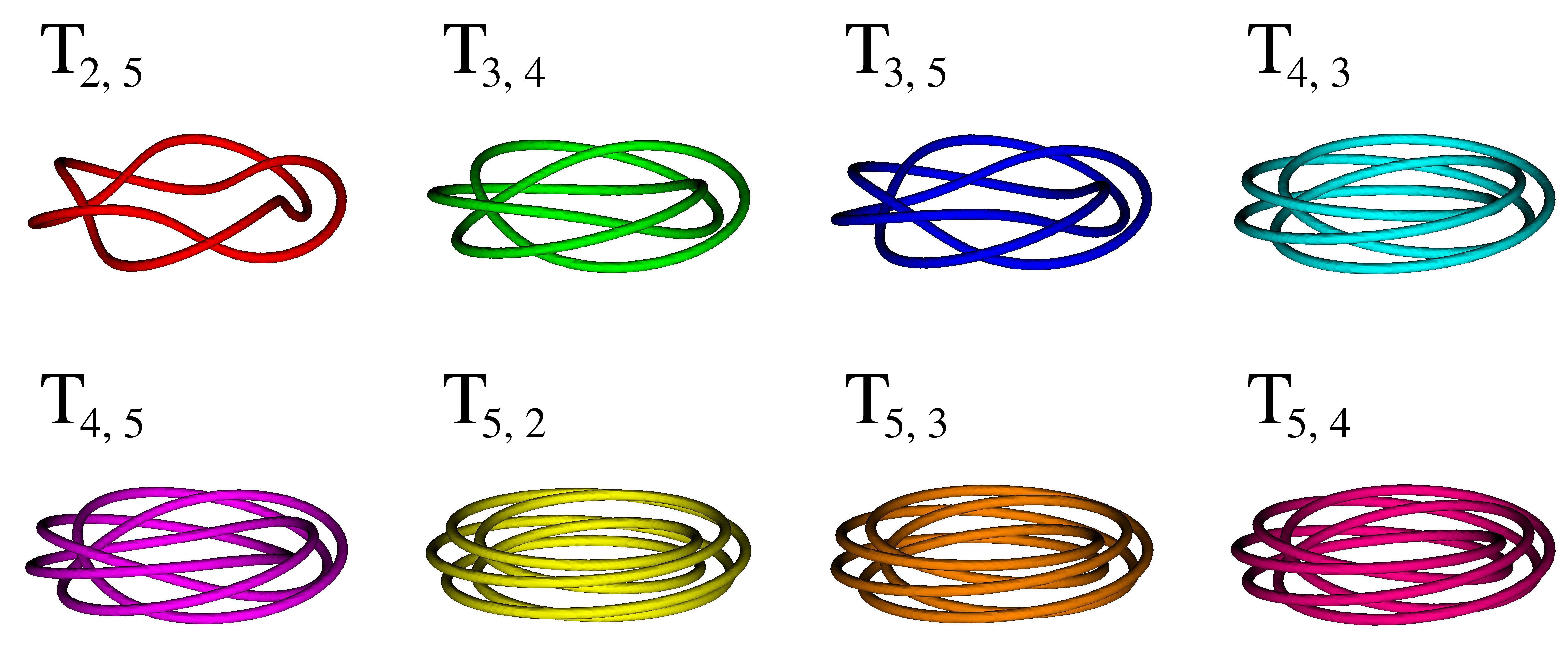}\hspace{2pc}%
\begin{minipage}[b]{0.9\linewidth}\caption{
Iso-surface density plots at $ \rho_{th}=0.2 \, \bar{\rho} $ showing the initial configuration of different quantum knots $ \mathcal{T}_{p, q} $ based on the wave-function (\ref{eq:Tpq}) having $ R_0=20 $ and $ R_1=4 $ (lengths are expressed in healing length units).
\label{fig:initialKnotAll}}
\end{minipage}
\end{figure}

Quantum vortices are easily detectable as points where the density field $ \rho(x, y, z) $ vanishes.
In a system where the mean spatial density over the entire volume is $ \bar{\rho} $ it is usually sufficient to plot the iso-surfaces corresponding to the threshold $ \rho_{th}=0.2 \, \bar{\rho} $ to show vortex defects and to filter out eventual sound density waves.
By using equation (\ref{eq:Tpq}) and setting $ R_0=20 $ and $ R_1=4 $ we can plot the iso-surfaces at $ \rho_{th} $ for various knots, as shown in Figure \ref{fig:initialKnotAll}, and use their wave-function as initial condition to study their dynamics.


\section{Vortex knot energy}
The above structures modeled by GPE represent collective excitations of the superfluid.
The non-dimensional GPE (\ref{eq:ndGPE}) has two conserved quantities which are the total number of bosons $ N $ and the total energy $ E $, resulting in the following integrals over the space volume $ V $:
\index[subject]{superfluid}
\begin{equation}
N=\int_V |\psi(\mathbf{r}, t)|^2 dV \quad \mbox{and} \quad 
E=\int_V \left( \frac{1}{2} |\nabla \psi(\mathbf{r}, t)|^2 + \frac{1}{4} |\psi(\mathbf{r}, t)|^4 \right) dV \, ,
\end{equation}
where $ dV=dxdydz $ and $ \mathbf{r}=(x, y, z) $.
Set the mean density $ \bar{\rho}=N/V $, one can identify the energy corresponding to an excitation given its wave-function.
It is well-know that in the absence of an external potential the ground state wave-function for the GPE (\ref{eq:ndGPE}) is the uniform density solution $ \psi_{gs}(x, y, z)=\sqrt{\bar{\rho}} \, \exp{ \left[-i (\bar{\rho}/2 \, t + \alpha_0 ) \right]} $, where $ \alpha_0 \in [0, 2\pi) $ is an overall phase that depends on the initial conditions.
Thus, the excitation energy of a particular structure with respect to the ground state is simply
\index[subject]{excitation energy}
\begin{equation}
\begin{split}
E_{exc} = & \int_V \left( \frac{1}{2} |\nabla \psi(\mathbf{r}, t)|^2 + \frac{1}{4} |\psi(\mathbf{r}, t)|^4 \right) dV 
- \int_V \left( \frac{1}{2} |\nabla \psi_{gs}(\mathbf{r}, t)|^2 + \frac{1}{4} |\psi_{gs}(\mathbf{r}, t)|^4 \right) dV \\
= & \int_V \left( \frac{1}{2} |\nabla \psi(\mathbf{r}, t)|^2 + \frac{1}{4} |\psi(\mathbf{r}, t)|^4 \right) dV - \frac{\bar{\rho}^2 \, V}{4} \, .
\label{eq:exc}
\end{split}
\end{equation}

\begin{figure}[!ht]
\begin{minipage}{0.45\linewidth}
\includegraphics[width=\linewidth]{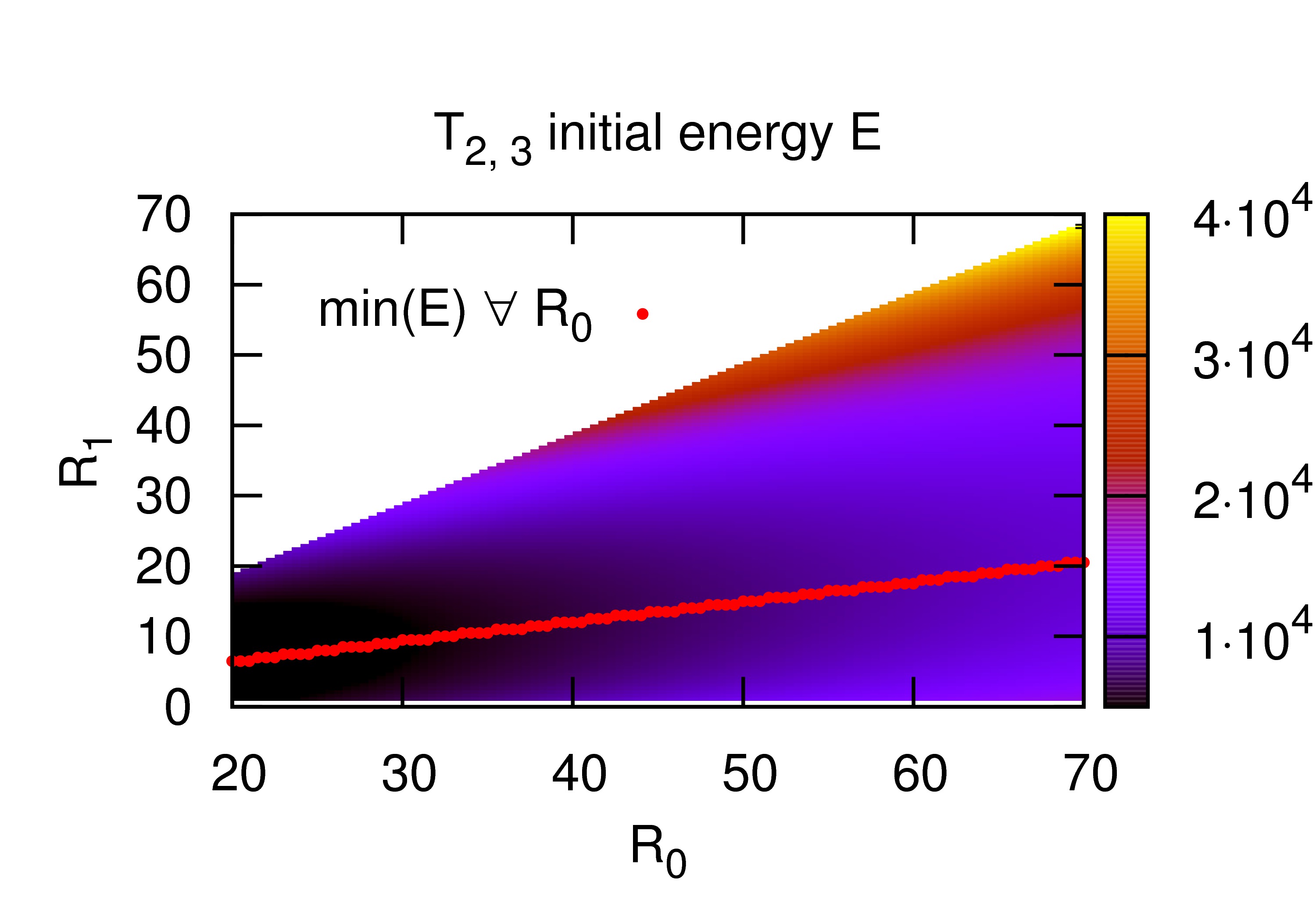}
\caption{Computed excitation energy (\ref{eq:exc}) for the quantum vortex knot $ \mathcal{T}_{2, 3} $ as a function of the torus radii $ R_0 $ and $ R_1 $.
The red points are the energy minima evaluated by fixing the toroidal radius $ R_0 $ (lengths are expressed in healing length units).
\index[subject]{excitation energy}
\label{fig:enT23}}
\end{minipage}
\hspace{0.08\linewidth}%
\begin{minipage}{0.45\linewidth}
\includegraphics[width=\linewidth]{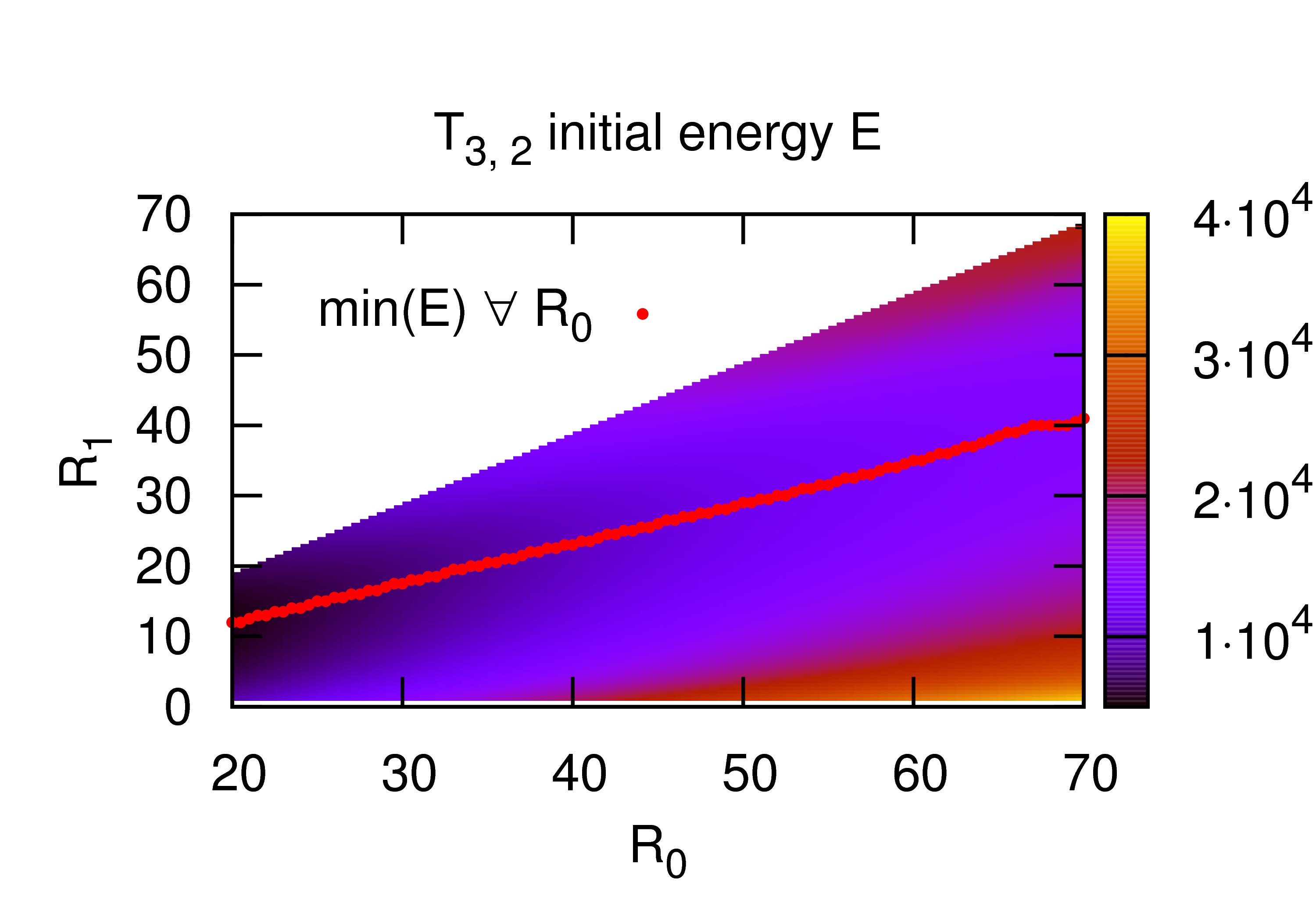}
\caption{Computed excitation energy (\ref{eq:exc}) for the quantum vortex knot $ \mathcal{T}_{3, 2} $ as a function of the torus radii $ R_0 $ and $ R_1 $.
The red points are the energy minima evaluated by fixing the toroidal radius $ R_0 $ (lengths are expressed in healing length units). 
\index[subject]{excitation energy}
\label{fig:enT32}}
\end{minipage} 
\end{figure}

\begin{figure}[!ht]
\includegraphics[width=0.45\linewidth]{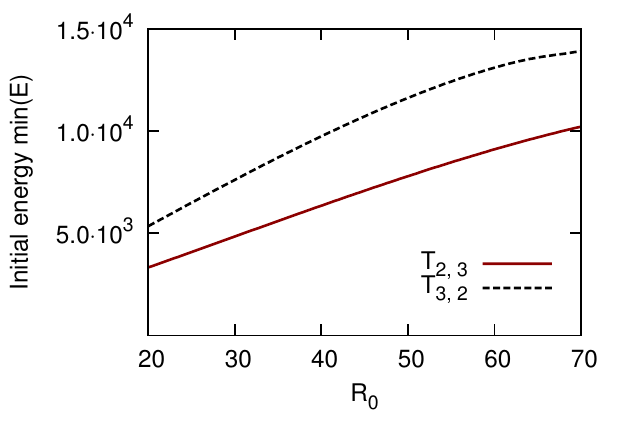}
\hspace{0.08\linewidth}%
\begin{minipage}[b]{0.45\linewidth}
\caption{Computed excitation energy minima with respect to $ R_0 $ in the range of second parameter $ R_1 \in [1; R_0-1] $ for the $ \mathcal{T}_{2, 3} $ and $ \mathcal{T}_{3, 2} $ knots - $ R_1 $ is in this interval in order to have always a properly defined torus (lengths are expressed in healing length units). 
\index[subject]{excitation energy}
\vspace{1.6cm}
\label{fig:enMin}}
\end{minipage}
\end{figure}

\index[subject]{excitation energy}
\index[subject]{trefoil knot}
One may then ask what is the excitation energy of a generic $ \mathcal{T}_{p, q} $ torus knot.  
We have addressed this problem numerically by considering a trefoil $ \mathcal{T}_{2, 3} $ knot and its dual $ \mathcal{T}_{3, 2} $ knot positioned at the centre of a finite cubic volume of side $ L=441 \, \Delta x $ (the numerical algorithms use a uniform grid spacing $ \Delta x = \xi/2 $ and 441 points are selected to increase efficiency of the cosine transforms).
Results obtained by considering $ R_0 $ and $ R_1 $ as parameters are presented in Figures \ref{fig:enT23} and \ref{fig:enT32} corresponding to the wave-function (\ref{eq:Tpq}) of the $ \mathcal{T}_{2, 3} $ and $ \mathcal{T}_{3, 2} $ torus knots respectively.
We analyzed the parameter space $ R_0 \in [20; 70] $ and $ R_1 \in [1; R_0-1] $ in order to have always well defined tori with $ R_1 < R_0 $ and knots always smaller than two thirds of the system size to limit boundary effects.
We observe that in both cases the energy is a monotonically function of the toroidal radius $ R_0 $.
We have evaluated the energy minima fixing $ R_0 $ and plotted as red points.
It is evident that in both cases the minima form a straight line in the parameter plane: our best fits give
\begin{equation}
\begin{split} 
& R_1^{(2, 3)}=(0.2816 \pm 0.0010)R_0 + (0.807 \pm 0.047) \\
& R_1^{(3, 2)}=(0.5811 \pm 0.0022)R_0 + (0.05 \pm 0.10) 
\end{split}
\label{eq:enFits}
\end{equation}
for the $ \mathcal{T}_{2, 3} $ and $ \mathcal{T}_{3, 2} $ cases respectively.
We plot the energy minimum values as a function of $ R_0 $ in Figure \ref{fig:enMin} in order to compare between the two cases.
For fix values of $ R_0 $, the energy of the $ \mathcal{T}_{2, 3} $ knot appears to be much less with respect to the $ \mathcal{T}_{3, 2} $ knot one: this is somehow expected as the former propagates much slower than the latter [\cite{PhysRevE.85.036306}].
\index[authors]{Proment, D.}
\index[authors]{Onorato, M.}
\index[authors]{Barenghi, C.F.}
One could for completeness obtain numerically the momentum-energy relations for different $ \mathcal{T}_{p, q} $ knots and compare them to the same relation for the perfect vortex rings [\cite{Jones:1982fk}] but is out of our present scope. 
\index[authors]{Jones, C.A.}
\index[authors]{Roberts, P.H.}





\section{Knot dynamics}
The easiest measurable quantities giving information about the knot dynamics are its centre of mass $ \mathbf{r}_{CM}(t)=\left[x_{CM}(t), y_{CM}(t), z_{CM}(t)\right] $ and its characteristic size $ \mathbf{D}=(D_x, D_y, D_z) $.
These can be defined as
\begin{equation}
\mathbf{r}_{CM}(t) = \frac{\int_V \mathbf{r} \, H(\rho_{th}-|\psi(\mathbf{r}, t)|^2) \, dV}{\int_V H(\rho_{th}-|\psi(\mathbf{r}, t)|^2) \, dV}\label{eq:CM}
\end{equation}
and
\begin{equation}
\mathbf{D}(t) = \max_{\mathbf{r},  \, \mathbf{r}' \, \in \, V} \left[ \mathbf{r} \, H(\rho_{th}-|\psi(\mathbf{r}, t)|^2)- \mathbf{r}' \, H(\rho_{th}-|\psi(\mathbf{r}', t)|^2)\right]
\end{equation}
respectively, being $ \mathbf{r}'=(x', y', z') $, $ H(\cdot) $ the Heaviside step function and $ \rho_{th} = 0.2 \, \bar{\rho} $.

\begin{figure}[!ht]
\begin{minipage}{0.45\linewidth}
\includegraphics[width=\linewidth]{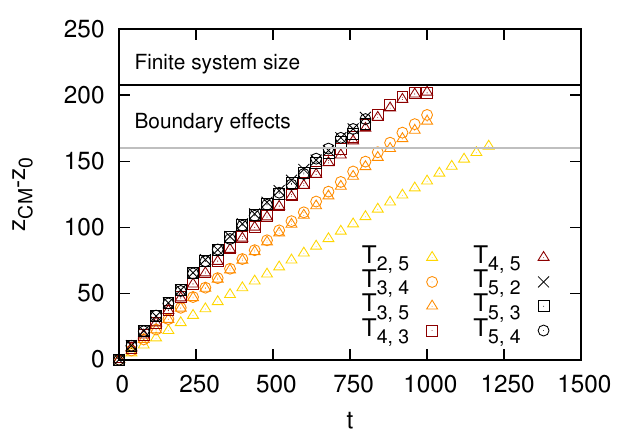}
\caption{Evolution of the displacement in the axisymmetric torus $ z $ direction of different $ \mathcal{T}_{p, q} $ knots (lengths are expressed in healing length units).
Knots having the same number of $ p $ twists are shown using the same colour, while the same symbol is used to underline the same number of $ q $ twists.
Estimated boundary and finite size effects are also shown.
\label{fig:zCM}}
\end{minipage}
\hspace{0.08\linewidth}%
\begin{minipage}{0.45\linewidth}
\includegraphics[width=\linewidth]{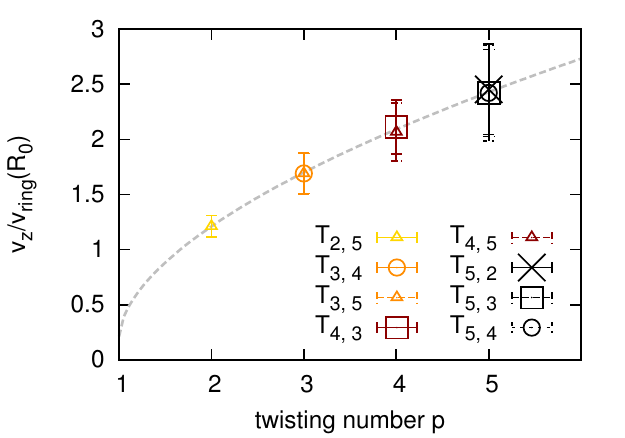}
\caption{Estimated $ z $ velocity component $ v_z $ normalised using $ v_{\mbox{ring}}(R_0) $ for different $ \mathcal{T}_{p, q} $ knots.
Knots having the same number of $ p $ twists are shown using the same colour, while the same symbol is used to underline the same number of $ q $ twists.
The grey dashed line is the best fit with the function $ f(p)=(p-1)^a+c $, with $ a=0.5747\pm0.0026 $ and $ c=0.210\pm0.011 $.
\label{fig:vz}}
\end{minipage} 
\end{figure}

Torus knots propagate essentially along the torus axis (here the $ z $ axis) twisting around it.
Thus, the translation can be quantified measuring the evolution of the centre of mass component $ z_{CM} $, which is plotted in Figure \ref{fig:zCM}, for different torus knots.
It appears evident that $ \mathcal{T}_{p, q} $ knots with equal twisting number $ p $ (here plotted using the same colour) have the same motion along the $ z $ axis; moreover, to a first approximation, this motion is linear.
The estimated velocity component $ v_z $ for different $ T_{p, q} $ knots is shown in Figure \ref{fig:vz}, normalised by the velocity of a vortex ring of radius $ R_0 $. Following \cite{donnelly1991}, the ring velocity is
\begin{equation}
v_{\mbox{ring}}(R)= \frac{\kappa}{4 \pi R} \left[\ln\left(\frac{8 R}{\xi}\right) - 0.615 \right] \, ,
\end{equation}
where we have set $ \xi=1 $ and $ \kappa=2\pi $ to be coherent with our non-dimensional GPE (\ref{eq:ndGPE}).
Confirming previous observations, it results that the torus knot translational velocity depends only on the twisting number $ p $, being for toroidal and poloidal radi $ R_0=20 $ and $ R_1=4 $ always greater than $ v_{\mbox{ring}}(R_0) $.
Moreover, we found that the behaviour of $ v_z/v_{\mbox{ring}}(R_0) $ with respect to the twisting number $ p $ is very well fitted by the function $ f(p)=(p-1)^a+c $, plotted with dashed grey line, with $ a=0.5747\pm0.0026 $ and $ c=0.210\pm0.011 $ coefficients.

\begin{figure}[!ht]
\includegraphics[width=0.45\linewidth]{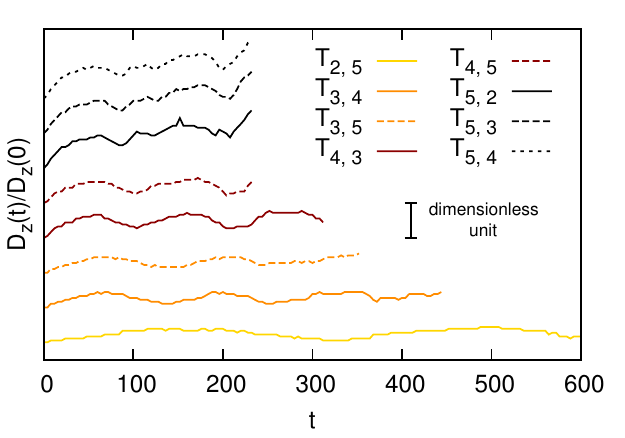}
\hspace{0.08\linewidth}%
\begin{minipage}[b]{0.45\linewidth}
\caption{Evolution of the characteristic size $ D_z $ rescaled with the corresponding initial size for different quantum vortex knots.
A dimensionless unit is also plotted in order to appreciate the fluctuations of around 25\% of each rescaled size.
\vspace{1.6cm}
\label{fig:sizeZ}}
\end{minipage}
\end{figure}

During the evolution the characteristic knot size visibly varies. 
In Figure \ref{fig:sizeZ} we plot the characteristic size component $ D_z $ for the different knots analysed, rescaled with their respective initial size. 
It appears that for every knots considered the rescaled size component $ D_z $ oscillates (around 25\%) with a characteristic frequency that seems to depend only on the twisting number $ p $.
We underline that especially for knots having high twisting number $ p $, $ \mathcal{T}_{5, q} $ for instance, the characteristic size $ D_z $ also grows in time.
\index[subject]{reconnection}





\section{Decaying mechanisms}
In \cite{PhysRevE.85.036306} we have shown that initial $ \mathcal{T}_{2, 3} $ and $ \mathcal{T}_{3, 2} $ torus knots having $ R_0=20 $ and $ R_1 \ge 8 $ do not preserve their shape during the evolution and decay after several reconnection events.
The decay product is a set of vortex rings which have different radii and carry Kelvin waves on them.
In general we observed that knots having twisting number $ p $ produce (at least) $ p $ rings, with a decay mechanism that seems to be peculiar to the initial topology. 
\index[authors]{Proment, D.}
\index[authors]{Onorato, M.}
\index[authors]{Barenghi, C.F.}
\index[subject]{Kelvin waves}
\index[subject]{reconnection}

We report here a mere observation on how the more topologically complicated $ \mathcal{T}_{p, q} $ torus knots shown in Figure \ref{fig:initialKnotAll} having $ R_0=20 $ and $ R_1=12 $ break (visit \url{http://youtu.be/whxqwdp_ogA} for a complete movie).
In all knots tested we observe always two phenomena which we believe to be universal.
First, the final products of the breaking event(s) are $ p $ vortex rings, which may eventually produce even more rings if their Kelvin waves have enough amplitude to trigger single vortex ring self-reconnections.
This is somehow expected, as, far enough from the vortex knot, the initial phase field plotted in a generic plane $ sOz $ is similar to the one having $ p $ vortex rings axisymetric to the $ z $ axis and sufficiently close together. 
Second, all self-reconnections which break up the initial $ \mathcal{T}_{p, q} $ knot by producing the final $ p $ rings, are always simultaneous (this may be a consequence of the symmetry of the initial condition) and $ q $ in number.
\index[subject]{Kelvin waves}
\index[subject]{reconnection}

\section{Open questions and future perspectives}
Many open questions remain on BEC quantum knot static and dynamical properties.
We will address in the following the more interesting ones and list what are the future perspectives in this field. 

Understanding the stability of knotted structures is probably the most difficult mathematical issue.
Many studies have been carried out using the local induction approximation of the Biot-Savart model and recently it has been shown that a $ \mathcal{T}_{p, q} $ torus knot with $ p > q > 1 $ are unstable, while in the case $ q > p $ it could be neutrally stable and quickly changes its knot type during the evolution [\cite{Calini:2011fk, Calini:2011uq}].
However, it was noted numerically that the full Biot-Savart model has a stabilization effect allowing unstable knots to travel for distances longer with respect to their characteristic size [\cite{ricca:1999, PhysRevE.82.026309}].
As far as we are aware, no mathematical studies on stability have ever been attempted using the GPE.
Is the $ \mathcal{T}_{p, q} $ torus knot for a particular set of $ (p, q) $ a Hamiltonian or shape-preserving structure?
\index[authors]{Calini, A.}
\index[authors]{Keith, S.F.}
\index[authors]{Lafortune, S.}
\index[authors]{Calini, A.}
\index[authors]{Ivey, T.}
\index[authors]{Ricca, R.L.}
\index[authors]{Samuels, D.C.}
\index[authors]{Barenghi, C.F.}
\index[authors]{Maggioni, F.}
\index[authors]{Alamri, S.}
\index[authors]{Barenghi, C.F.}
\index[authors]{Ricca, R.L.}
\index[subject]{Biot-Savart model}
\index[subject]{local induction approximation}
\index[subject]{stability}

Regarding non shape-preserving knots, it has been observed that breaking mechanisms seem to follow specific universal rules.
At present, only torus knots were investigated but many other examples can be studied including much more complicated topologies like other knots and links.
What are the breaking mechanisms for a generic initial topology?
What are the intermediate breaking products and the type and number of final ones?
\index[subject]{breaking mechanisms}

Finally, one would like to understand what are the possibilities to find such kind of topologically non-trivial vortex structures in reality.
Can they occur naturally during (or after) a turbulent decay?
Otherwise, what are the experimental techniques in superfluids to create them in a controlled and reproducible way?
What kind of detecting techniques may be used to detect knotted and linked structures?
\index[subject]{superfluid}

Hopefully, in the following years some of these questions will find an answer and new advances in this field may help to solve open problems in related disciplines like classical and magneto-hydrodynamics. 
For instance, during the final drafting of this work, two different manuscripts addressing helicity conservation during reconnection events have appeared online as preprints [\cite{2014arXiv1403.8121B, 2014arXiv1404.6513S}].
\index[authors]{Baggaley, A.W.}
\index[authors]{Scheeler, M.W.} 
\index[authors]{Kleckner, D.} 
\index[authors]{Proment, D.} 
\index[authors]{Kindlmann, G.L.} 
\index[authors]{Irvine, W.T.M.}  
\index[subject]{helicity}

\section*{Acknowledgements}
The authors acknowledge the Isaac Newton Institute for Mathematical Sciences in Cambridge and the Julian Schwinger Foundation for supporting there the {\it Topological Dynamics in the Physical and Biological Sciences} program.
D. P. is grateful to A. Baggaley, M. Dennis, D. Kleckner, W. Irvine, D. Maestrini, R. Ricca, and H. Salman for fruitful suggestions and discussions.
The research presented in this paper was carried out on the High Performance Computing Cluster supported by the Research and Specialist Computing Support service at the University of East Anglia.


\bibliographystyle{jfm2}

\bibliography{TOD-bibliography}

\begin{thebibliography}{24}
\expandafter\ifx\csname natexlab\endcsname\relax\def\natexlab#1{#1}\fi

\bibitem[Abo-Shaeer {\em et~al.\/}(2001)Abo-Shaeer, Raman, Vogels \&
  Ketterle]{Abo-Shaeer20042001}
{\sc Abo-Shaeer, J.~R., Raman, C., Vogels, J.~M. \& Ketterle, W.} 2001
  {Observation of Vortex Lattices in Bose-Einstein Condensates}. {\em
  Science\/} {\bf 292}~(5516), 476--479.

\bibitem[Anderson {\em et~al.\/}(2001)Anderson, Haljan, Regal, Feder, Collins,
  Clark \& Cornell]{PhysRevLett.86.2926}
{\sc Anderson, B.~P., Haljan, P.~C., Regal, C.~A., Feder, D.~L., Collins,
  L.~A., Clark, C.~W. \& Cornell, E.~A.} 2001 {Watching Dark Solitons Decay
  into Vortex Rings in a Bose-Einstein Condensate}. {\em Phys. Rev. Lett.\/}
  {\bf 86}~(14), 2926--2929.

\bibitem[Anderson {\em et~al.\/}(1995)Anderson, Ensher, Matthews, Wieman \&
  Cornell]{anderson1995observation}
{\sc Anderson, M.~H., Ensher, J.~R., Matthews, M.~R., Wieman, C.~E. \& Cornell,
  E.~A.} 1995 {Observation of Bose-Einstein condensation in a dilute atomic
  vapor}. {\em Science\/} {\bf 269}~(5221), 198--201.

\bibitem[{Baggaley}(2014)]{2014arXiv1403.8121B}
{\sc {Baggaley}, A.~W.} 2014 {Helicity transfer during quantised vortex
  reconnection}. {\em ArXiv e-prints\/} {\bf {}}, 1403.8121.

\bibitem[Berloff(2004)]{berloff:2004}
{\sc Berloff, N.} 2004 {Pad{\'e} approximations of solitary wave solutions of
  the Gross-Pitaevskii equation}. {\em Journal of Physics A: Mathematical and
  General\/} {\bf 37}, 1617--1632(16).

\bibitem[Bewley {\em et~al.\/}(2008)Bewley, Paoletti, Sreenivasan \&
  Lathrop]{bewley2008characterization}
{\sc Bewley, G.~P., Paoletti, M.~S., Sreenivasan, K.~R. \& Lathrop, D.~P.} 2008
  Characterization of reconnecting vortices in superfluid helium. {\em
  Proceedings of the National Academy of Sciences\/} {\bf 105}~(37),
  13707--13710.

\bibitem[Calini \& Ivey(2011)]{Calini:2011fk}
{\sc Calini, A. \& Ivey, T.} 2011 Stability of small-amplitude torus knot
  solutions of the localized induction approximation. {\em Journal of Physics
  A: Mathematical and Theoretical\/} {\bf 44}~(33).

\bibitem[Calini {\em et~al.\/}(2011)Calini, Keith \& Lafortune]{Calini:2011uq}
{\sc Calini, A., Keith, S.~F. \& Lafortune, S.} 2011 Squared eigenfunctions and
  linear stability properties of closed vortex filaments. {\em Nonlinearity\/}
  {\bf 24}~(12).

\bibitem[Davis {\em et~al.\/}(1995)Davis, Mewes, Andrews, Van~Druten, Durfee,
  Kurn \& Ketterle]{davis1995bose}
{\sc Davis, K., Mewes, M.-O., Andrews, M.~v., Van~Druten, N., Durfee, D., Kurn,
  D. \& Ketterle, W.} 1995 {Bose-Einstein condensation in a gas of sodium
  atoms}. {\em Physical Review Letters\/} {\bf 75}~(22), 3969.

\bibitem[Demokritov {\em et~al.\/}(2006)Demokritov, Demidov, Dzyapko, Melkov,
  Serga, Hillebrands \& Slavin]{demokritov2006bose}
{\sc Demokritov, S., Demidov, V., Dzyapko, O., Melkov, G., Serga, A.,
  Hillebrands, B. \& Slavin, A.} 2006 {Bose--Einstein condensation of
  quasi-equilibrium magnons at room temperature under pumping}. {\em Nature\/}
  {\bf 443}~(7110), 430--433.

\bibitem[Donnelly(1991)]{donnelly1991}
{\sc Donnelly, R.} 1991 {\bf {Quantized vortices in Helium II}}. , vol.~3.
  Cambridge Univ Pr.

\bibitem[Jones \& Roberts(1982)]{Jones:1982fk}
{\sc Jones, C.~A. \& Roberts, P.~H.} 1982 Motions in a bose condensate. iv.
  axisymmetric solitary waves. {\em Journal of Physics A: Mathematical and
  General\/} {\bf 15}~(8), 2599.

\bibitem[Kasprzak {\em et~al.\/}(2006)Kasprzak, Richard, Kundermann, Baas,
  Jeambrun, Keeling, Marchetti, Szyma{\'n}ska, Andr{\'e}, Staehli, Savona,
  Littlewood, Deveaud \& Dang]{kasprzak2006bose}
{\sc Kasprzak, J., Richard, M., Kundermann, S., Baas, A., Jeambrun, P.,
  Keeling, J., Marchetti, F., Szyma{\'n}ska, M.~H., Andr{\'e}, R., Staehli,
  J.~L., Savona, V., Littlewood, P.~B., Deveaud, B. \& Dang, L.~S.} 2006
  {Bose--Einstein condensation of exciton polaritons}. {\em Nature\/} {\bf
  443}~(7110), 409--414.

\bibitem[Klaers {\em et~al.\/}(2010)Klaers, Schmitt, Vewinger \&
  Weitz]{klaers2010bose}
{\sc Klaers, J., Schmitt, J., Vewinger, F. \& Weitz, M.} 2010 {Bose-Einstein
  condensation of photons in an optical microcavity}. {\em Nature\/} {\bf
  468}~(7323), 545--548.

\bibitem[Kleckner \& Irvine(2013)]{kleckner2013creation}
{\sc Kleckner, D. \& Irvine, W. T.~M.} 2013 Creation and dynamics of knotted
  vortices. {\em Nature Physics\/} {\bf 9}~(4), 253--258.

\bibitem[Madison {\em et~al.\/}(2000)Madison, Chevy, Wohlleben \&
  Dalibard]{PhysRevLett.84.806}
{\sc Madison, K.~W., Chevy, F., Wohlleben, W. \& Dalibard, J.} 2000 {Vortex
  Formation in a Stirred Bose-Einstein Condensate}. {\em Phys. Rev. Lett.\/}
  {\bf 84}~(5), 806--809.

\bibitem[Maggioni {\em et~al.\/}(2010)Maggioni, Alamri, Barenghi \&
  Ricca]{PhysRevE.82.026309}
{\sc Maggioni, F., Alamri, S., Barenghi, C.~F. \& Ricca, R.~L.} 2010 Velocity,
  energy, and helicity of vortex knots and unknots. {\em Phys. Rev. E\/} {\bf
  82}, 026309.

\bibitem[Matthews {\em et~al.\/}(1999)Matthews, Anderson, Haljan, Hall, Wieman
  \& Cornell]{PhysRevLett.83.2498}
{\sc Matthews, M.~R., Anderson, B.~P., Haljan, P.~C., Hall, D.~S., Wieman,
  C.~E. \& Cornell, E.~A.} 1999 {Vortices in a Bose-Einstein Condensate}. {\em
  Phys. Rev. Lett.\/} {\bf 83}~(13), 2498--2501.

\bibitem[Neely {\em et~al.\/}(2010)Neely, Samson, Bradley, Davis \&
  Anderson]{PhysRevLett.104.160401}
{\sc Neely, T.~W., Samson, E.~C., Bradley, A.~S., Davis, M.~J. \& Anderson,
  B.~P.} 2010 {Observation of Vortex Dipoles in an Oblate Bose-Einstein
  Condensate}. {\em Phys. Rev. Lett.\/} {\bf 104}~(16), 160401.

\bibitem[Pitaevskii \& Stringari(2003)]{pitaevskii2003bose}
{\sc Pitaevskii, L. \& Stringari, S.} 2003 {\bf {Bose-Einstein Condensation}}.
  , vol. 116. Oxford University Press, USA.

\bibitem[Proment {\em et~al.\/}(2012)Proment, Onorato \&
  Barenghi]{PhysRevE.85.036306}
{\sc Proment, D., Onorato, M. \& Barenghi, C.~F.} 2012 {Vortex knots in a
  Bose-Einstein condensate}. {\em Phys. Rev. E\/} {\bf 85}, 036306.

\bibitem[Ricca {\em et~al.\/}(1999)Ricca, Samuels \& Barenghi]{ricca:1999}
{\sc Ricca, R., Samuels, D.~C. \& Barenghi, C.~F.} 1999 Evolution of vortex
  knots. {\em J. Fluid Mech.\/} {\bf 391}, 29--44.

\bibitem[{Scheeler} {\em et~al.\/}(2014){Scheeler}, {Kleckner}, {Proment},
  {Kindlmann} \& {Irvine}]{2014arXiv1404.6513S}
{\sc {Scheeler}, M.~W., {Kleckner}, D., {Proment}, D., {Kindlmann}, G.~L. \&
  {Irvine}, W.~T.~M.} 2014 {Helicity conservation in topology-changing
  reconnections: the flow of linking and coiling across scales}. {\em ArXiv
  e-prints\/} {\bf {}}, 1404.6513.

\bibitem[Weiler {\em et~al.\/}(2008)Weiler, Neely, Scherer, Bradley, Davis \&
  Anderson]{Weiler:2008fk}
{\sc Weiler, C.~N., Neely, T.~W., Scherer, D.~R., Bradley, A.~S., Davis, M.~J.
  \& Anderson, B.~P.} 2008 {Spontaneous vortices in the formation of
  Bose-Einstein condensates}. {\em Nature\/} {\bf 455}~(7215), 948--951.

\end{thebibliography}

\end{document}